\documentclass[prb,twocolumn,superscriptaddress,float,aps]{revtex4-2}
\usepackage{bm,color,amsmath,amssymb,mathrsfs,latexsym,graphicx,psfrag,tabularx}

\usepackage[hidelinks,colorlinks,linkcolor=blue,
citecolor=blue,urlcolor=blue]{hyperref}

\newcommand{\beq}{\begin{equation}}
\newcommand{\eneq}{\end{equation}}

\def\qq{\mathbf{q}}
\def\kk{\mathbf{k}}
\def\KK{\mathbf{K}}
\def\DKK{\Delta\mathbf{K}}
\def\pp{\mathbf{p}}
\def\RR{\mathbf{R}}
\def\tt{\mathbf{t}}
\def\rr{\mathbf{r}}
\def\GG{\mathbf{G}}
\def\QQ{\mathbf{Q}}
\def\aa{\mathbf{a}}
\def\bb{\mathbf{b}}
\def\uu{\mathbf{u}}

\def\KK{\mathbf{K}}
\def\qq{\mathbf{q}}
\def\pp{\mathbf{p}}
\def\pp{\mathbf{p}}
\def\GG{\mathbf{G}}
\def\QQ{\mathbf{Q}}
\def\RR{\mathbf{R}}
\def\tt{\mathbf{t}}
\def\dd{\mathbf{d}}
\def\aa{\mathbf{a}}
\def\bb{\mathbf{b}}
\def\ee{\epsilon}
\def\CC{\mathcal{P}}

\def\BZ{{\rm BZ}}
\def\mS{{\mathcal{S}}}

\def\BZ{{\rm BZ}}

\def\spin{{\varsigma}}

\def\spin{{\varsigma}}

\def\hH{{ \hat{H} }}
\def\hrho{ \hat{\rho} }
\def\hg{\hat{g}}
\def\hS{\hat{S}}

\def\mG{{\mathcal{G}}}

\def\UC{{\hat{\Theta}}}
\def\UF{{\hat{\Sigma}}}

\def\mK{{\mathcal{K}}}
\def\pr{\prime}
\def\mJ{{\mathcal{J}}}
\def\mK{{\mathcal{K}}}

\def\ie{{\it i.e.},\ }
\def\eg{{\it e.g.}\ }
\def\ea{{\it et al.}}

\usepackage{etoolbox}
\makeatletter
\patchcmd{\@maketitle}{\@author}{\@author\show\@thanks}{}{}
\makeatother

\usepackage{dsfont}

\begin{document}

\title{Perfect Nonreciprocal Axion-polaritons}

\author{Abhinava Chatterjee}
\email{abhinava.chatterjee@psu.edu}
\affiliation{%
   Department of Physics, The Pennsylvania State University, University Park, Pennsylvania 16802, USA
}
\author{Chao-Xing Liu}
\affiliation{%
   Department of Physics, The Pennsylvania State University, University Park, Pennsylvania 16802, USA
}
\affiliation{Center for Theory of Emergent Quantum Matter, The Pennsylvania State University, University Park, Pennsylvania 16802, USA}

\begin{abstract} 
Dynamical axion describes a massive axion quasiparticle that arises from fluctuations of the magnetic order parameter. These fluctuations couple to electromagnetic fields, forming axion-polaritons --- collective modes that are physically magnon-polaritons. We show that under appropriate static external electric and magnetic fields, axion-polaritons acquire a nonreciprocal dispersion,  $\omega(\textbf{k}) \neq \omega(-\textbf{k})$, arising intrinsically from direction-dependent axion-photon coupling. Strikingly, we identify a regime of \textit{perfect nonreciprocity} in which the photon propagating in one direction is completely decoupled from the axion while the counter-propagating photon hybridizes strongly. Furthermore, the nonreciprocal dispersion manifests as an optical isolator --- a device transmitting light preferentially in one direction. Our results establish nonreciprocal axion-polaritons as a powerful probe of axion quasiparticles and suggest experimentally accessible routes for their detection.
\end{abstract}

\date{\today}


\maketitle

\textit{Introduction} ---
Axion electrodynamics describes a topological electromagnetic response characterized by a $\theta_0 {\bf E \cdot B}$ term in the effective action \cite{qi2008topological}, where $\theta_0$ is the static axion field quantized to $0$ or $\pi$ by time-reversal (TRS) or inversion symmetry \cite{qi2008topological,wilczek1987two}. Originally introduced to address the strong $CP$-violation problem in high-energy physics \cite{peccei1977cp,peccei1977constraints,weinberg1978new,wilczek1978problem}, the axion field is now recognized as an emergent degree of freedom in condensed matter systems \cite{wilczek1987two,qi2008topological,essin2009magnetoelectric,nenno2020axion,sekine2021axion}. When both inversion and TRS (or the effective $S$ symmetry) are broken, the axion field is no longer topologically constrained to $0$ or  
$\pi$ and becomes dynamical, acquiring its own equation of motion as a propagating degree of freedom \cite{li2010dynamical,wang2016dynamical,sekine2016chiral,liu2021making,sekine2021axion,nenno2020axion}. Fluctuations of the magnetic order parameter give rise to a massive axion quasiparticle, $\theta$, coupled to electromagnetic fields through $\mathcal{L}_{\theta_0} \sim (\theta_0 + \theta) {\bf E \cdot B} $. Signatures of dynamical axion fields have been observed in Weyl and Dirac semimetals
\cite{wang2013chiral,jzbr-b2p8,gooth2019axionic,wieder2020axionic,you2016response,curtis2023finite,mottola2024axions} and a variety of theoretical detection schemes have been proposed \cite{smith2026theory,boyanovsky2025probing,liebman2025multiphoton,liebman2025exciton,gao2025detecting,gomes2025probing}. Beyond magnetic systems, axion quasiparticles associated with lattice degrees of freedom have also been proposed  \cite{lhachemi2024phononic,bernabeu2025axionic}. On the material front, van der Waals MnBi$_2$Te$_5$-related compounds have been predicted to host the axion quasiparticle  \cite{zhang2020large,li2020Mn2,cao2021growth}. More recently, Kerr rotation measurements in two-dimensional MnBi$_2$Te$_4$ have interpreted antiferromagnetic magnon excitations as signature of axion quasiparticles \cite{qiu2025observation} - representing the most compelling experimental evidence to date. As with any nascent experimental observation, independent signatures from complementary physical probes are essential to unambiguously establish the axion quasiparticle.

The coupling between the dynamical axion and electromagnetic fields leads to the formation of axion polaritons \cite{li2010dynamical,maiani1986effects,raffelt1988mixing,cameron1993search}, including surface modes residing within the polaritonic gap \cite{zhu2022axionic} and axionic dark modes in cavity geometries \cite{xiao2023short}. Related axionic phenomena include electric-field induced instabilities due to axionic screening \cite{ooguri2012instability,taguchi2018electromagnetic,imaeda2019axion} and nonreciprocal thermal emission \cite{zhao2020axion}. Despite these advances, nonreciprocal axion-polaritons, namely propagating axion-photon hybrid modes characterized by $\omega({\bf k}) \neq \omega(-{\bf k}) $, have not been identified in either condensed matter or high-energy physics. Prior work on axion-polaritons has relied solely on a static magnetic field \cite{li2010dynamical,zhu2022axionic,xiao2023short}, which breaks TRS but preserves inversion symmetry, and hence cannot produce $\omega({\bf k}) \neq \omega(-{\bf k}) $. Similarly, high-energy physics experiments probing axion-photon coupling - including PVLAS and CAST - employ magnetic fields alone \cite{maiani1986effects,raffelt1988mixing,cameron1993search,
zavattini2006experimental,cast2017new}, which is insufficient to generate directional asymmetry in the dispersion. Achieving nonreciprocal axion-polaritons requires simultaneously breaking both inversion and TRS through combined electric and magnetic fields - a regime not previously explored. 

Optical isolators, devices that transmit light only in one direction, are 
indispensable components in optical networks and quantum 
information processing \cite{jalas2013and}. Conventional 
isolators rely on the magneto-optical Faraday effect \cite{jalas2013and}, chiral 
edge states in magneto-optical photonic crystals 
\cite{wang2009observation,el2015optical}, Weyl 
semimetal-based isolators \cite{chistyakov2023tunable,
karki2019toward}, spin-orbit locking using 
nanophotonic waveguides in chiral quantum optics 
\cite{lodahl2017chiral}, and time-modulation schemes that break 
TRS through active driving \cite{sounas2017non,
fang2012realizing}. In the 
context of magnon-polariton systems, nonreciprocal 
transmission has been demonstrated in lossy magnon 
cavities \cite{zhang2017observation}, meta-molecular 
systems \cite{mita2025ultrastrongly}, and twisted 
magnetic structures \cite{peng2021twist}. However, in 
all these platforms, nonreciprocity relies on either 
external biasing, engineered dissipation, active 
modulation, or nanoscale confinement. The possibility 
of optical isolation arising intrinsically from 
direction-dependent axion-photon coupling in a bulk 
material tunable by static fields has not been proposed.

\begin{figure}
\centering
\includegraphics[width=\columnwidth]{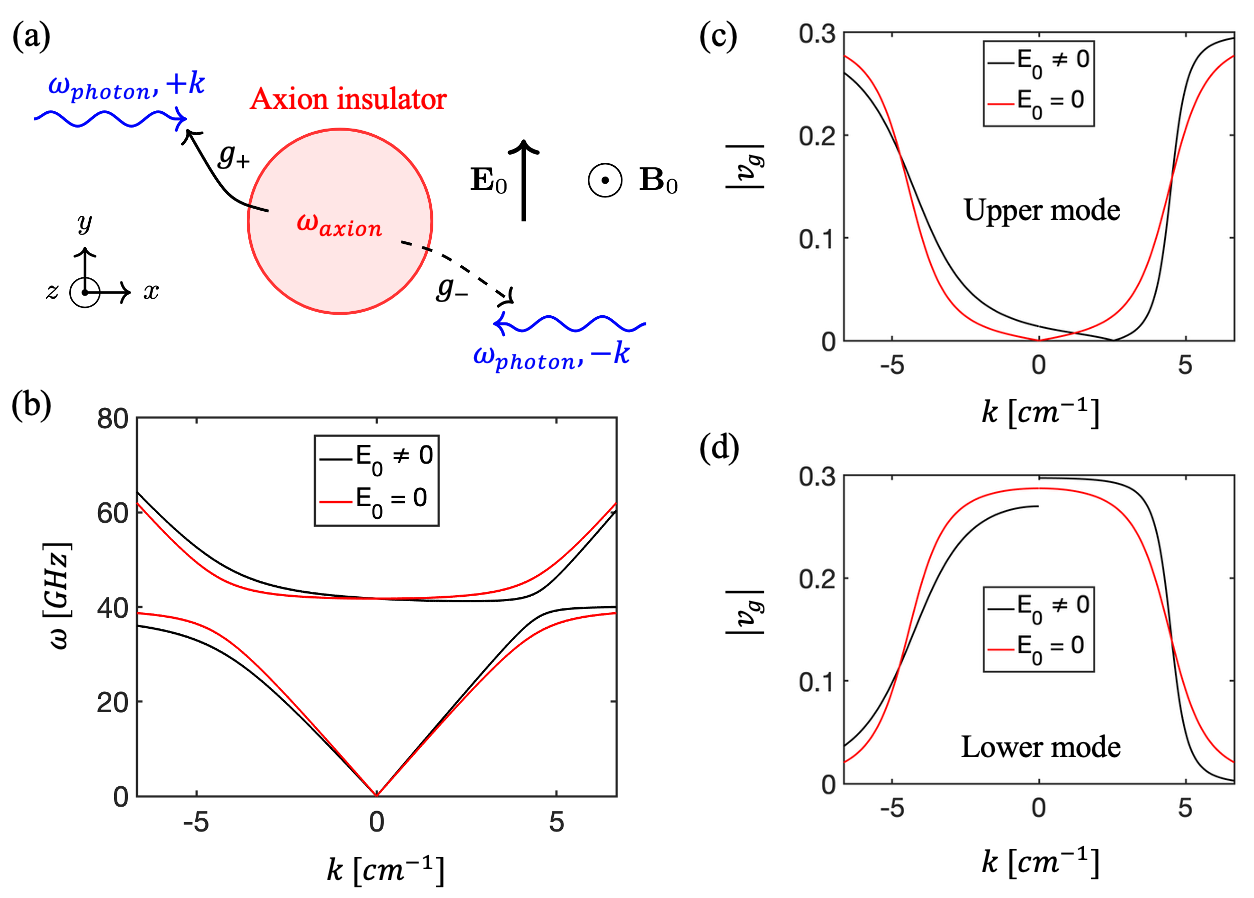}
\caption{ (a) The dynamical axion mode interacting with photons. In the presence of both external electric and magnetic field $E_0, B_0$ along $y$ and $z$ directions respectively, the photons from the right $(\omega_{\text{photon}}, -k)$ and from the left $(\omega_{\text{photon}}, +k)$ couple to the axion mode with different coupling strengths $g_+ \neq g_-$ (defined in the section \textit{perfect nonreciprocity}). (b) The frequency of the axion-polariton modes in the presence ($E_0 \neq 0$ - black) and absence of an external electric field ($E_0 = 0$ - red) as a function of momentum $k$. The momentum axis is in units of cm$^{-1}$, obtained using the conversion $1$ GHz $= 0.0333$ cm$^{-1}$. The magnitude of the group velocity $|v_g|$ of the (c) upper and (d) lower polariton modes when $E_0 \neq 0$ (black) and $E_0 = 0$ (red). We use $\mathcal{E} = 2$ GHz, $\mathcal{B} = 12$ GHz, $m=40$ GHz, $v = 0.005$ and $c'=0.3$. 
}
\label{fig:NRAP}
\end{figure}

In this work, we propose nonreciprocal axion-polaritons, establishing axion-photon coupling as a fundamentally new route to nonreciprocity. We study a dynamical axion insulator coupled to the electromagnetic field in the presence of static external electric and magnetic fields, as illustrated in Fig. \ref{fig:NRAP}(a). We show that the resulting axion-polaritons exhibit intrinsic nonreciprocity, characterized by asymmetric dispersions $\omega_{{\bf k}} \neq \omega_{-{\bf k}}$. Microscopically, this effect arises from the direction-dependent axion-photon coupling strengths $g_+ \neq g_-$ (Fig. \ref{fig:NRAP}(a)), with counter-propagating photon modes hybridizing with the axion quasiparticle with unequal amplitudes. Remarkably, the nonreciprocity can be tuned by external electromagnetic fields, enabling regimes of \textit{perfect nonreciprocity}, in which the right-moving photon mode is completely decoupled from the axion mode, while the left-moving photon mode hybridizes strongly. Furthermore, we 
compute the transmission scattering matrix and show that this direction-dependent coupling manifests itself directly as an 
optical isolator in which the right-moving photon propagates freely through the slab, while the left-moving photon, 
hybridizing with the axion mode, is exponentially attenuated.

\begin{figure*}
\includegraphics[width= \textwidth]{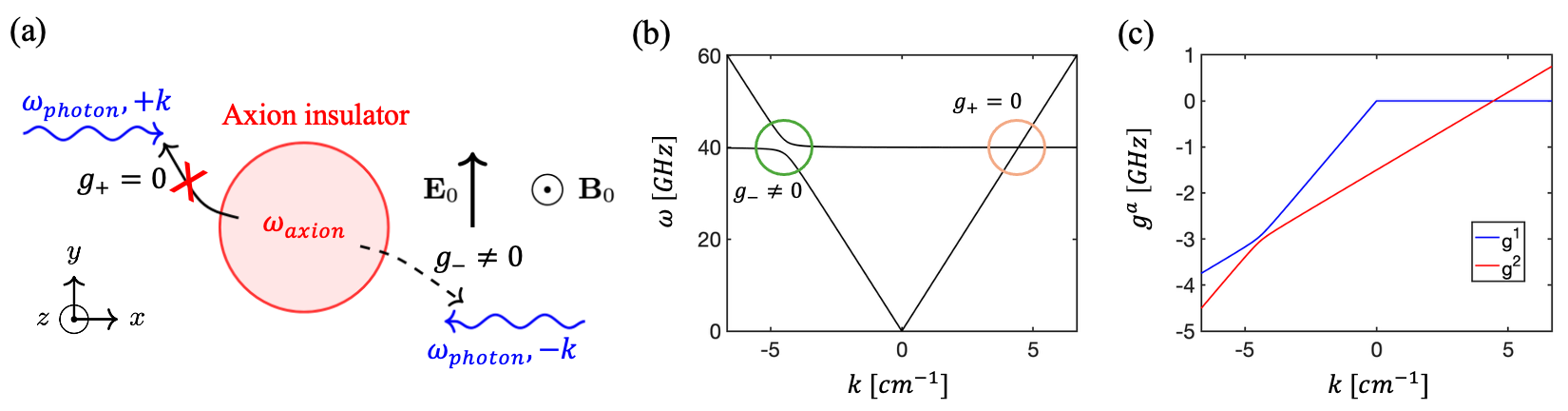}
\caption{ (a) \textit{Perfect nonreciprocity} at critical electromagnetic fields ${\bf E_0^c, B_0^c}$ given by $E_0^c = c' B_0^c$. (b) The axion-polariton dispersion at the critical external field. (c) The axion-photon coupling strengths $g^a(k)$ as a function of momentum $k$ at critical electromagnetic fields ${\bf E_0^c, B_0^c}$. We use $\mathcal{E} = c' \mathcal{B}$, $\mathcal{B} = 1.5$ GHz, $m=40$ GHz, $v = 0.005$ and $c'=0.3$.
}
\label{fig:Perfect}
\end{figure*}

\textit{Nonreciprocal axion-polaritons} --- Dynamical axion insulators are described by the Lagrangian 
\cite{li2010dynamical}
\begin{align}\label{eq_M:AILagrangan}
    \mathcal{L} &= \mathcal{L}_{EM} + \mathcal{L}_{\theta_0} + \mathcal{L}_{\theta} \nonumber \\
    &= \frac{1}{8 \pi} \left(  \epsilon {\bf E}^2 - \frac{1}{\mu} {\bf B}^2\right) + \frac{\alpha}{4 \pi^2} \left(  \theta_0 + \theta\right) {\bf E \cdot B} \nonumber \\ &+ J \left( \left( \partial_t \theta \right)^2 - \left( v_i \partial_i \theta\right)^2 - m^2 \theta^2 \right),
\end{align}
where $\epsilon$ is the electric permittivity, $\mu$ is the magnetic permeability, ${\bf E}$ and ${\bf B}$ are the electromagnetic fields inside the axion insulator, $\alpha$ is the fine structure constant, $\theta_0$ is the static field quantized as $0$ or $\pi$, $\theta$ is the dynamical axion field, $J$ is the magnon stiffness, $v_i$ is the axion velocity along $i=x,y,z$ and $m$ is the mass of the dynamical axion field. The electromagnetic field couples to the axion field through $\mathcal{L}_{\theta_0}$ yielding hybrid axion-photon modes, dubbed axion-polaritons \cite{li2010dynamical}. In the presence of both external electromagnetic fields ${\bf E_0, B_0} $, the coupled equation of motion of the axion and the photon becomes $ M \Phi = 0$ where $\Phi = ({\bf A},\theta)^T$ in the Coulomb gauge $\nabla \cdot {\bf A} = 0$ with $A_0 = 0$ and the matrix $M$ is given by
\begin{widetext}
\begin{equation} \label{eq_M:MMatrix}M = 
   \begin{pmatrix}
       \Box_A & 0 & 0 & -\frac{\alpha}{4 \pi^2} \left( B_0^x \partial_t + E_0^z \partial_y - E_0^y \partial_z \right) \\
       0 & \Box_A & 0 &   -\frac{\alpha}{4 \pi^2} \left( B_0^y \partial_t + E_0^x \partial_z- E_0^z \partial_x \right) \\ 0 & 0 & \Box_A & -\frac{\alpha}{4 \pi^2} \left( B_0^z \partial_t + E_0^y \partial_x - E_0^x \partial_y \right) \\ 
   \frac{\alpha}{4 \pi^2} \left( B_0^x \partial_t + E_0^z \partial_y - E_0^y \partial_z \right)    & \frac{\alpha}{4 \pi^2} \left( B_0^y \partial_t + E_0^x \partial_z- E_0^z \partial_x \right) & \frac{\alpha}{4 \pi^2} \left( B_0^z \partial_t + E_0^y \partial_x - E_0^x \partial_y \right) & \Box_\theta 
   \end{pmatrix}
\end{equation}
\end{widetext}
where $\Box_A = \frac{\epsilon}{4 \pi} \left(\partial_t^2 - c'^2 \partial_i^2 \right) $, $\Box_\theta = 2 J \left( \partial_t^2 - v_i^2 \partial_i^2 + m^2 \right) $, $c'$ is the photon velocity inside the dynamical axion insulator, $c' = 1/\sqrt{\mu \epsilon}$, and the photon velocity in vacuum is taken to be unity, $c=1$. We consider a static external electric field along $y$-axis, ${\bf E_0} \parallel \Hat{y}$ and a static external magnetic field along $z$-axis, ${\bf B_0} \parallel \Hat{z}$, as shown in Fig. \ref{fig:NRAP}(a). We take the ansatz $({\bf A}, \theta) = (\Tilde{{\bf A}}, \Tilde{\theta}) e^{i (k x - \omega t)}$ where ${\bf k} = k \Hat{x}$. From Eq.(\ref{eq_M:MMatrix}), $A_{x,y}$ decouple from $A_{z}$ and $\theta$. Furthermore, $A_x = 0$ in the Coulomb gauge. The coupled equation of motion of $A_z$ and $\theta$ is given by
\begin{equation} \label{eq_M:EOM}
    \begin{pmatrix}
        \frac{\epsilon}{4 \pi} \left( -\omega^2 + c'^2 k^2 \right)  & -\frac{i \alpha}{4 \pi^2} \left( E_0^y k - \omega B_0^z \right) \\ \frac{i \alpha}{4 \pi^2} \left( E_0^y k - \omega B_0^z \right) & 2 J \left( -\omega^2 + v^2 k^2 + m^2 \right) 
    \end{pmatrix} \begin{pmatrix}
        \Tilde{A}_z \\ \Tilde{\theta}
    \end{pmatrix} = 0
\end{equation}
The frequency $\omega$ satisfies
\begin{align}
    \omega^4  &- \Big[ \left( v^2 + c'^2\right) k^2 + m^2 + \mathcal{B}^2 \Big] \omega^2 +  2 \mathcal{E B} k \omega \nonumber \\ +& \Big[ c'^2 k^2 \left( v^2 k^2 + m^2 \right) - \mathcal{E}^2 k^2 \Big] =0 \label{eq_M:weqn}
\end{align}
where $\mathcal{E} = \frac{\alpha E_0^y}{\sqrt{\epsilon J (2 \pi)^3}}$, $\mathcal{B} = \frac{\alpha B_0^z}{\sqrt{\epsilon J (2 \pi)^3}}$. In the limit $\mathcal{E} = 0, \mathcal{B} \neq 0$ in Eq.(\ref{eq_M:weqn}), we recover the reciprocal axion-polariton dispersion in Ref. \cite{li2010dynamical}, while the case with $\mathcal{E} \neq 0, \mathcal{B} = 0$ recovers the electric-field driven instability of Ref. \cite{ooguri2012instability}. Crucially, neither limit alone can produce nonreciprocal dispersion. The key to nonreciprocity lies in the term $2 \mathcal{EB} k \omega$ that is linear in both $\omega$ and $k$ in Eq.(\ref{eq_M:weqn}), which requires the simultaneous presence of both $\mathcal{E}$ and $\mathcal{B}$: the static electric field breaks inversion symmetry while the static magnetic field $\mathcal{B}$ breaks TRS. In Fig. \ref{fig:NRAP}(b), we plot the frequency of the axion-polaritons. When $\mathcal{E} = 0$, axion-polaritons are formed \cite{li2010dynamical} with reciprocal dispersion $\omega_{{\bf k}} = \omega_{-{\bf k}} $, shown by the red curve in Fig. \ref{fig:NRAP}(b). When the external electric field is turned on, $\mathcal{E} \neq 0$, the axion-polaritons display nonreciprocity $\omega_{{\bf k}} \neq \omega_{-{\bf k}} $, for a range of momenta, as shown by the black lines in Fig. \ref{fig:NRAP}(b). At small momenta ${\bf k} \ll m/c' \sim 4.5$ cm$^{-1}$, the axion-polaritons are almost decoupled modes due to large energy separation between the axion and the photon modes. The higher energy polariton mode is nearly a pure axion mode $\omega_1({\bf k} \rightarrow 0) = \sqrt{  v^2 k^2 + m^2 } $, and the lower energy polariton mode is nearly a pure photon $\omega_2({\bf k} \rightarrow 0) = c' |k| $. Since the axion velocity $v$ is much smaller than the speed of light in the medium $c'$, the dispersion of the axion mode is almost negligible compared to that of the pure photon. For an intermediate range of momenta ${\bf k} \sim m/c'$, the energy separation is reduced, leading to an avoided crossing and strong hybridization - the defining feature of axion-polariton formation. For large momenta ${\bf k} \gg m/c'$, the modes are again nearly decoupled, similar to their behavior at small momenta. However, due to the exchanged mode character near the avoided crossing at intermediate ${\bf k}$, the higher energy mode is now photon-like and the lower energy mode is axion-like. The nonreciprocity of axion-polaritons is directly reflected in the asymmetry of the absolute value of the group velocity $|v_g ({\bf k})| = |\partial_{{\bf k}} \omega({\bf k})|$ between positive and negative momentum directions. In Fig. \ref{fig:NRAP}
(c), we plot $|v_g|$ for the upper axion-polariton branch. When $\mathcal{E} = 0$ (red), $|v_g|$ is symmetric with respect to $k$, as expected for a reciprocal system. At $k=0$, the upper branch is flat and the group velocity vanishes. With increasing $|k|$, the upper branch acquires photon character and the group velocity approaches the photon velocity $c' \sim 0.3$. When $\mathcal{E} \neq 0 $ (black), a clear asymmetry emerges: the group velocity vanishes at finite $k \sim 2.5$ cm$^{-1}$ rather than at $k=0$, and the two sides $\pm k$ approach $c'$ at different rates, directly reflecting the nonreciprocal dispersion. Similarly, in Fig. \ref{fig:NRAP}
(d), we plot $|v_g|$ for the lower axion-polariton branch. When $\mathcal{E} = 0$ (red), $|v_g|$ is again symmetric. At $k=0$, the lower branch is linear and the group velocity is approximately the photon velocity $c' \sim 0.3$. With increasing $|k|$, the lower branch acquires axion character and the group velocity decreases toward zero, reflecting the nearly flat dispersion of the axion mode. When $\mathcal{E} \neq 0 $ (black), the asymmetry is strikingly visible even at $k=0$: the group velocities for $-k$ and $+k$ are unequal, a direct consequence of the nonreciprocal dispersion. At asymptotic $k$, the group velocity approaches zero from both sides but at different rates.

\textit{Perfect nonreciprocity} --- We now turn to a special limit of our setup: when the external electromagnetic fields satisfy $ |{\bf E_0^c}| = c' |{\bf B_0^c}| $, the nonreciprocity in the axion-photon coupling becomes perfect. As illustrated in Fig. \ref{fig:Perfect}
(a), the right-moving photon is completely decoupled from the axion mode while the left-moving photon hybridizes strongly --- a direct consequence of the directional selectivity inherent to nonreciprocal axion-polaritons. To quantify this, we define the direction-dependent coupling strength for a particular axion-polariton branch $a$ as $g^a_\pm = g^a(\pm |k|)$ where
\begin{equation}\label{eq:geqn}
    g^a(k) = \frac{1}{m}\left( E_0^y k - \omega_a(k) B_0^z \right) ,
\end{equation}
where $a=1,2$. Since $g^a(k)$ depends on the mode frequency $\omega_a(k)$, the two axion-polariton branches generally yield different coupling strengths $g^1(k) \neq g^2(k)$ at the same momentum $k$. At the perfect nonreciprocity condition $\mathcal{E} = c' \mathcal{B}$, $\omega = c' k$ is an exact solution of Eq.(\ref{eq_M:weqn}), while $\omega=c' |k|$ is not. This establishes that the right-moving photon propagates freely, completely unaffected by the axion mode, while the left-moving photon hybridizes strongly. In Fig. \ref{fig:Perfect}(b), we plot the axion-polariton dispersion in the perfect nonreciprocity limit. For $k>0$, the photon mode is decoupled from the axion mode (orange circle) and the photon dispersion is unchanged. For $k<0$, both branches are strongly renormalized by the coupling (green circle). The microscopic origin of this asymmetry is revealed in Fig. \ref{fig:Perfect}(c), where we plot the coupling strengths $g^a(k)$ for both axion-polariton branches as a function of $k$. One branch maintains $g^1=0$ (blue) for all $k>0$ - corresponding to the completely decoupled right-moving photon mode, whereas both branches exhibit $g^{1,2} \neq 0$ for $k<0$, confirming strong hybridization in the left-moving sector. The physical origin of perfect nonreciprocity is most transparent when Eq.(\ref{eq_M:EOM}) is recast as Maxwell's equations with axion-induced source terms:
\begin{align}
    & \nabla \cdot {\bf E} =  - \frac{\alpha}{ \pi } \nabla \theta \cdot {\bf B_0} \label{eq_M:ChargeSource} \\
     & \nabla \times {\bf B} = \frac{1}{c'^2} \frac{\partial {\bf E}}{\partial t}  - \frac{\alpha}{\pi } \left( \Dot{\theta} {\bf B_0} + \nabla \theta \times {\bf E_0} \right) .\label{eq_M:CurrentSource} 
\end{align}
The charge source in Eq.(\ref{eq_M:ChargeSource}) vanishes identically since $\nabla \theta \sim \Hat{x} \perp {\bf B_0}$. Furthermore, the Coulomb gauge implies $\nabla \cdot {\bf A} =0$ and Eq.(\ref{eq_M:ChargeSource}) is trivially satisfied. The axion-induced current source experienced by the photon in Eq.(\ref{eq_M:CurrentSource}) takes the form $ j_{\text{ax}} \sim \Dot{\theta} {\bf B_0} + \nabla \theta \times {\bf E_0} \sim  - \omega B_{0}^z + k E_{0}^y  $, which vanishes when $E_0^y = c' B_0^z $ and $\omega = +c' k$. Therefore, the axion mode provides no source term for the right-moving photon, which propagates as a free photon. It is important to note, however, that perfect nonreciprocity does not imply complete decoupling of the right-moving sector. While the right-moving photon experiences no axion-induced source current, the axion mode in the right-moving sector remains coupled to the photon, as evidenced by $g^2 \neq 0$ (red) for the axion branch at $k > 0$ in Fig.~\ref{fig:Perfect}(c). This one-way coupling — where the photon 
acts as a source for the axion but not vice versa — is a hallmark 
feature of perfect nonreciprocity and distinguishes it from trivial 
decoupling. Such asymmetric coupling is a novel feature of nonreciprocal 
axion-polaritons. 

\begin{figure}
\centering
\includegraphics[width=\columnwidth]{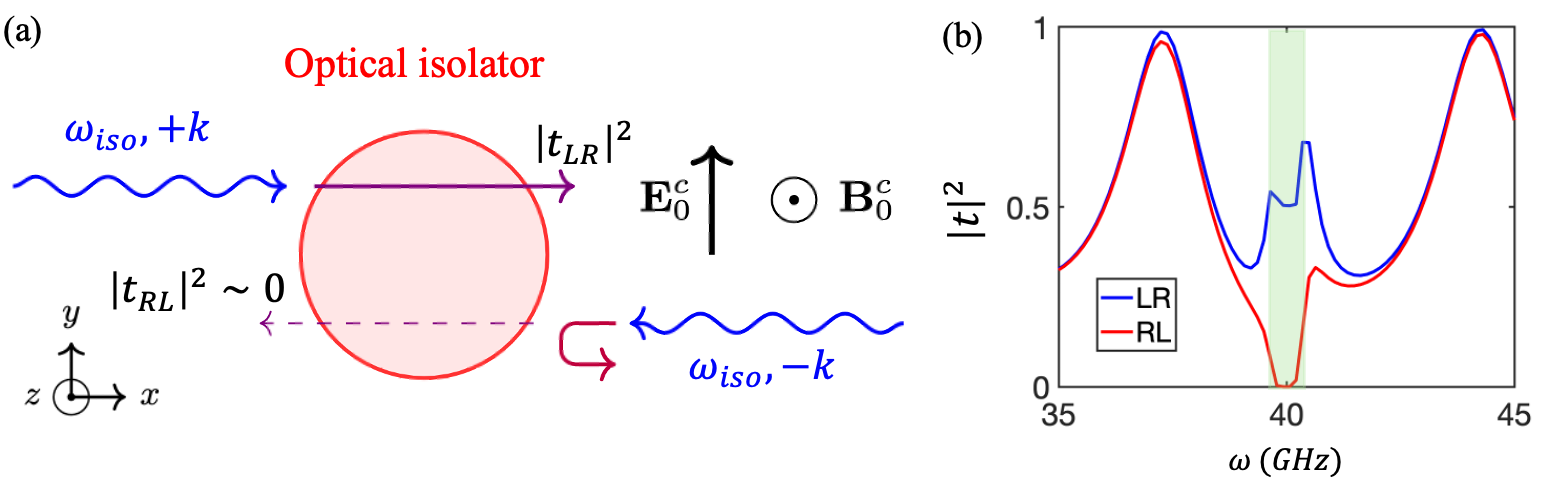}
\caption{ (a) The axion insulator as an \textit{optical isolator}. (b) The transmissivities $|t_{LR}|^2$ and $|t_{RL}|^2$ as a function of photon frequency $\omega$. We use $\mathcal{E} = c' \mathcal{B}$ , $\mathcal{B} = 1.5$ GHz, $m=40$ GHz, $v = 0$ and $c'=0.3$. 
}
\label{fig:isolator}
\end{figure}

\textit{Optical isolator} --- The nonreciprocal 
dispersion has a direct and experimentally accessible 
manifestation as an optical isolator (Fig. \ref{fig:isolator}(a)). We introduce a 
small dissipative term in the axion dynamics, 
$\sim \Gamma\partial_t\theta$, which models the finite 
lifetime of the axion quasiparticle (magnons), and compute the complex transmission coefficients $t_{LR}$ and $t_{RL}$ 
(see SM \cite{SM}), where $LR$ ($RL$) denotes 
left-to-right (right-to-left) transmission. The 
transmissivities $|t_{LR}|^2$ and $|t_{RL}|^2$ at 
perfect nonreciprocity $\mathcal{E} = c'\mathcal{B}$ are 
plotted in Fig.~\ref{fig:isolator}(b). We find that 
$|t_{LR}| \neq |t_{RL}|$ near the avoided crossing 
($\omega \sim 40$ GHz), with the right-to-left 
transmission strongly suppressed, while the two 
transmissivities are nearly equal away from the avoided 
crossing. This asymmetry has a transparent physical origin rooted 
in the nonreciprocal dispersion. At perfect 
nonreciprocity, the wavevector of the right photon mode $k_+ = \omega/c'$ is purely real regardless of $\Gamma$, which is a direct consequence of perfect nonreciprocity: the right-moving photon is completely decoupled from the lossy axion mode and propagates freely through the slab, giving high $|t_{LR}|$, reduced below unity only by Fresnel reflection due to the velocity mismatch 
$c \neq c'$ at the interfaces. In contrast, the wavevector of the left-moving photon acquires a finite imaginary part due to $\Gamma$, since 
the left-moving photon hybridizes strongly with the dissipative axion mode. As a result, photon energy is absorbed by the axion, causing exponential attenuation and strongly suppressed $|t_{RL}|$ near the avoided crossing. 
Crucially, $|t_{LR}| = |t_{RL}|$ identically when $\mathcal{E} = 0$ (see SM \cite{SM}), confirming that optical isolation requires the simultaneous breaking of 
both inversion and TRS through combined $\mathcal{E}$ 
and $\mathcal{B}$ fields. This establishes a 
fundamentally new mechanism for optical isolation arising intrinsically 
from the direction-dependent axion-photon coupling in a dynamical axion 
insulator.

\textit{Discussion and Conclusion} --- We have demonstrated the emergence of nonreciprocal axion-polaritons in a dynamical axion insulator subjected to mutually perpendicular static electric and magnetic fields. The nonreciprocal dispersion arises from the simultaneous breaking of inversion and TRS, and is characterized by direction-dependent group velocities for both polariton branches. We identified a special regime, termed \textit{perfect nonreciprocity} in which the right-moving photon is completely decoupled from the axion mode while the left-moving photon hybridizes strongly, realizing a one-way coupling. We further showed the realization of an optical isolator using this intrinsic direction-dependent axion-photon coupling mechanism. The nonreciprocal polariton modes are experimentally accessible using standard direction-dependent spectroscopic measurements and vector network analyzers (VNA), which give direct access to both 
amplitude and phase of $t_{LR}$ and $t_{RL}$ \cite{yao2025nonreciprocal, 
mita2025ultrastrongly}. We estimate the realistic static electromagnetic field strength required for perfect nonreciprocity. Using $J(2\pi)^3 = \alpha^2(1\,\text{T}/604.5\,\text{GHz})^2$\cite{ooguri2012instability,li2010dynamical}, we estimate that for $\mathcal{B} = 1.5$ GHz, the corresponding external fields are $E_0^y \sim 10^5 $ V/m and $B_0^z \sim 8$ mT. A static magnetic field of $\sim 8$ mT is readily achievable in laboratory settings. For a sample width of a few microns, a bias of $ \sim 1$ V suffices. The required static electric field can be further reduced by tuning the system close to the topological phase transition of the dynamical axion insulator \cite{shikin2023topological}. The axion mass $m=40$ GHz used in our calculations is motivated by recent Kerr rotation measurements in MnBi$_2$Te$_4$ \cite{qiu2025observation}, which identified antiferromagnetic magnon excitations as signatures of the axion quasiparticle at GHz frequencies. MnBi$_2$Te$_4$ and related van der Waals compounds \cite{zhang2020large,li2020Mn2,cao2021growth} thus provide compelling platforms for realizing nonreciprocal axion-polaritons. Our results apply broadly to antiferromagnetic insulators with magnetoelectric coupling and well-defined magnon modes. The magnetoelectric antiferromagnet Cr$_2$O$_3$ \cite{dzyaloshinskii1960magneto,astrov1960magnetoelectric,folen1961anisotropy} hosts magnon resonances at $\sim 0.16,0.24$ THz \cite{li2020spin}, making it a natural platform for nonreciprocal magnon-polaritons. Its dielectric breakdown field of hundreds of MV/m \cite{ashida2015isothermal,sun2017local,vu2020bulk} far exceeds the fields required here. 

The identification of nonreciprocal axion-polaritons and perfect nonreciprocity opens new directions for future exploration. It would be intriguing to investigate nonreciprocal slow light phenomena in dynamical axion insulators, analogous to studies in cavity magnonic systems \cite{yao2025nonreciprocal}. Moreover, nonreciprocal axion-polariton responses may offer enhanced sensitivity for dark matter axion detection \cite{marsh2019proposal,mitridate2020detectability,marsh2023axion}, suggesting a new avenue beyond conventional high-energy physics searches. 

\textit{Acknowledgments} --- The authors thank Yinming Shao and Mikael C. Rechtsman for helpful discussions. A.C. and C.-X.L. acknowledge support from NSF DMR-2241327 and the ONR Award (N000142412133).

\bibliographystyle{apsrev4-2}
\bibliography{ref}

\clearpage
\onecolumngrid    





\clearpage
\onecolumngrid

\section*{Supplementary material for ``Perfect Nonreciprocal Axion-polaritons"}

\setcounter{section}{0}
\setcounter{figure}{0}
\setcounter{table}{0}
\setcounter{equation}{0}

\renewcommand{\thefigure}{S\arabic{figure}}
\renewcommand{\thesection}{S\arabic{section}}
\renewcommand{\thetable}{S\arabic{table}}

\numberwithin{equation}{section}
\newcommand{\eqsref}[1]{Eq(S\ref{#1})}

\def\qq{\mathbf{q}}
\def\kk{\mathbf{k}}
\def\KK{\mathbf{K}}
\def\DKK{\Delta\mathbf{K}}
\def\pp{\mathbf{p}}
\def\RR{\mathbf{R}}
\def\tt{\mathbf{t}}
\def\rr{\mathbf{r}}
\def\GG{\mathbf{G}}
\def\QQ{\mathbf{Q}}
\def\aa{\mathbf{a}}
\def\bb{\mathbf{b}}
\def\uu{\mathbf{u}}
\def\AA{\mathbf{A}}
\def\ff{\mathbf{f}}
\def\mm{\mathbf{m}}

\def\KK{\mathbf{K}}
\def\qq{\mathbf{q}}
\def\pp{\mathbf{p}}
\def\pp{\mathbf{p}}
\def\GG{\mathbf{G}}
\def\QQ{\mathbf{Q}}
\def\RR{\mathbf{R}}
\def\tt{\mathbf{t}}
\def\dd{\mathbf{d}}
\def\aa{\mathbf{a}}
\def\bb{\mathbf{b}}
\def\ee{\epsilon}
\def\CC{\mathcal{P}}
\def\UU{\mathbf{U}}

\def\BZ{{\rm BZ}}
\def\mS{{\mathcal{S}}}

\def\spin{{\varsigma}}

\def\hH{{ \hat{H} }}
\def\hrho{ \hat{\rho} }
\def\hg{\hat{g}}
\def\hS{\hat{S}}

\def\mG{{\mathcal{G}}}

\def\UC{{\hat{\Theta}}}
\def\UF{{\hat{\Sigma}}}

\def\mK{{\mathcal{K}}}
\def\pr{\prime}
\def\mJ{{\mathcal{J}}}
\def\mK{{\mathcal{K}}}

\def\ie{{\it i.e.},\ }
\def\eg{{\it e.g.}\ }
\def\ea{{\it et al.}}


%
%


\pagenumbering{arabic}

\section{Transmission and reflection coefficients}

\subsection{Setup}

We consider a dynamical axion insulator slab of thickness 
$L$ occupying $0 < x < L$, with vacuum on both sides as shown in Fig. \ref{fig:Transmission}. Since $v << c'$, we take $v \rightarrow 0$. The axion equation of motion inside the slab becomes
\begin{equation}\label{eq_SM:axion_slaved}
    2J(m^2 - \omega^2 - i \Gamma \omega)\tilde{\theta} = 
    \frac{i\alpha}{4\pi^2}(E_0^yk - \omega B_0^z)
    \tilde{A}_z,
\end{equation}
where we have dropped the spatial gradient term 
$v^2\partial_x^2\theta$. We introduced a dissipation term $i \Gamma \omega$. Solving for $\tilde{\theta}$, we have
\begin{equation}\label{eq_SM:theta_slaved}
    \tilde{\theta} = -\frac{i\Lambda_1
    (E_0^yk - \omega B_0^z)}{m^2 - \omega^2 - i \Gamma \omega}
    \tilde{A}_z,
\end{equation}
where $\Lambda_1 = \alpha/8\pi^2J$. The axion field is completely determined by the photon field — it is not 
an independent degree of freedom in this limit.

\subsection{Effective photon dispersion}

\begin{figure*}
\includegraphics[width= \textwidth]{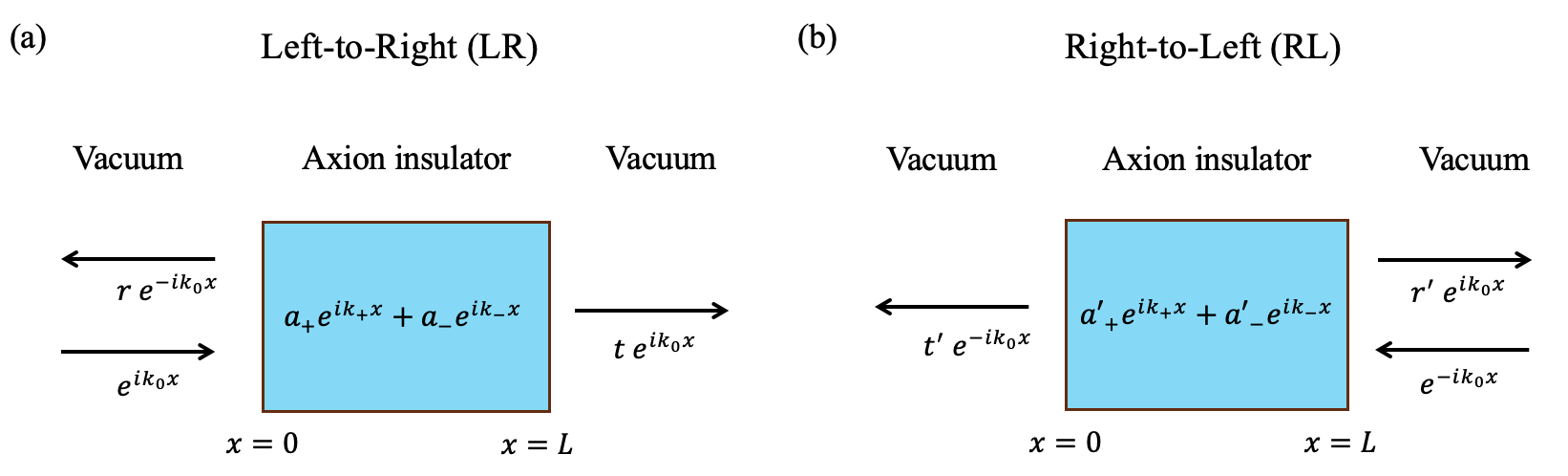}
\caption{ (a) The scattering setup for the left-to-right (LR) photon transmission and (b) the scattering setup for the right-to-left (RL) photon transmission.
}
\label{fig:Transmission}
\end{figure*}

Substituting Eq.~(\ref{eq_SM:theta_slaved}) into the 
photon equation of motion inside the slab , we obtain
\begin{equation}
    \frac{\epsilon}{4\pi}(-\omega^2 + c'^2k^2)
    \tilde{A}_z = \frac{i\alpha}{4\pi^2}
    (E_0^yk - \omega B_0^z)\tilde{\theta},
\end{equation}
and using Eq.~(\ref{eq_SM:theta_slaved}), the 
effective photon dispersion is
\begin{equation}\label{eq_SM:dispersion}
    \Big[ c'^2 \left( m^2 - \omega^2 - i \Gamma \omega \right) - \mathcal{E}^2 \Big] k^2 + 2 \omega \mathcal{EB}k + \Big[ \omega^4 + i \Gamma \omega^3 - \omega^2 \left( m^2 + \mathcal{B}^2 \right)  \Big] = 0.
\end{equation}
Eq.~(\ref{eq_SM:dispersion}) is quadratic in $k$, giving two solutions $k_+$ and $k_-$ for a given frequency $\omega$, given by
\begin{equation} \label{eq_SM:kpm1}
    k_\pm = \frac{-\omega \mathcal{EB} \pm c' \omega \sqrt{\left( \omega^2 + i \Gamma \omega \right)^2 + \left( \mathcal{E}^2/c'^2 - \mathcal{B}^2 - 2 m^2  \right) \left( \omega^2 + i \Gamma \omega \right)  + m^2 \left( m^2 + \mathcal{B}^2 - \mathcal{E}^2/c'^2 \right) }}{c'^2 \left( m^2 - \omega^2 - i \Gamma \omega \right) - \mathcal{E}^2} .
\end{equation}
The photon mode inside the axion insulator is a linear combination of $e^{i k_\pm x}$, as shown in Fig. \ref{fig:Transmission}. For general $\mathcal{E,B}$ values, $k_- \neq - k_+$, which implies that reciprocity is not guaranteed. When $\mathcal{E} = 0$, Eq.(\ref{eq_SM:kpm1}) becomes 
\begin{equation} \label{eq_SM:kpme0}
    k_\pm = \pm \frac{\omega}{c'} \sqrt{ \frac{\mathcal{B}^2 + m^2 - \omega^2 - i \Gamma \omega}{m^2 - \omega^2 - i \Gamma \omega} }.
\end{equation}
Setting $\Gamma = 0$ in Eq.(\ref{eq_SM:kpme0}), we find 
\begin{equation} \label{eq_SM:kpme0gamma0}
    k_\pm = \pm \frac{\omega}{c'} \sqrt{ \frac{\mathcal{B}^2 + m^2 - \omega^2 }{m^2 - \omega^2 } },
\end{equation}
and we recover the polaritonic gap when $m < \omega < \sqrt{m^2 + \mathcal{B}^2} $ as both $k_\pm$ become purely imaginary. 

Since $k_+ = - k_-$ (Eqs. \ref{eq_SM:kpme0},\ref{eq_SM:kpme0gamma0}), the transmission is expected to be reciprocal. In the perfect nonreciprocity limit $\mathcal{E} = c' \mathcal{B}$, Eq. (\ref{eq_SM:kpm1}) becomes
\begin{equation} \label{eq_SM:kpm2}
    k_+ = \frac{\omega}{c'}, \quad k_- = -\frac{\omega}{c'} \Big[ \frac{ \omega^2 + i \Gamma \omega - m^2 -\mathcal{B}^2}{\omega^2 + i \Gamma \omega - m^2  +\mathcal{B}^2} \Big] .
\end{equation}
From Eq. (\ref{eq_SM:kpm2}), it is clear that $k_+$ is the momentum of the decoupled photon mode shown in Fig. 2(b) of the main text, whereas $k_-$ is the momentum of the strongly coupled axion-polariton mode. Each mode has an 
associated axion field $\tilde{\theta}_\pm$ given by 
Eq.~(\ref{eq_SM:theta_slaved}) evaluated at $k = k_\pm$.

\subsection{Boundary conditions}

We derive the boundary conditions by integrating 
Eqs. (5) and (6) of the main text across the interfaces at 
$x=0,L$. Since ${\bf B}_0 \parallel \hat{z}$ and 
$\nabla\theta \parallel \hat{x}$, we have 
$\nabla\theta \cdot {\bf B}_0 = 0$, so 
Eq. (5) of the main text gives
\begin{equation}
    \nabla \cdot {\bf E} = 0,
\end{equation}
which implies continuity of the tangential electric 
field and hence continuity of $A_z$. At $x=0$,
\begin{equation}\label{eq_SM:Az_continuous}
    A_z(0^-) = A_z(0^+). 
\end{equation}
Integrating Eq. (6) of the main text across 
$x=0$ from $0^-$ to $0^+$, we have
\begin{equation}
    \int_{0^-}^{0^+}(\nabla\times{\bf B})_z\,dx = 
    \frac{1}{c'^2}\int_{0^-}^{0^+}\dot{E}_z\,dx 
    - \frac{\alpha}{\pi}\int_{0^-}^{0^+}
    \left(\dot{\theta}B_0^z + \partial_x\theta\, 
    E_0^y\right)dx.
\end{equation}
The first term on the right vanishes as 
$0^+-0^- \to 0$ since $\dot{E}_z$ is finite. Similarly, the 
$\dot{\theta}B_0^z$ term also vanishes and we obtain
\begin{equation}
    -\partial_xA_z(0^+) + 
    \partial_xA_z(0^-) = 
    -\frac{\alpha E_0^y}{\pi}
    \int_{0^-}^{0^+}\partial_x\theta\,dx = 
    -\frac{\alpha E_0^y}{\pi}
    \theta|_{0^-}^{0^+}.
\end{equation}
Since $\theta = 0$ outside the slab, we have
\begin{equation}\label{eq_SM:dAz_jump}
    \partial_xA_z(0^+) - 
    \partial_xA_z(0^-) = 
    \frac{\Lambda_2}{c'^2}   E_0^y
    \theta(0^+),
\end{equation}
where $\Lambda_2 = \alpha/\pi \epsilon$. The same procedure at $x=L$ gives
\begin{align}
    A_z(L^-) &= A_z(L^+) \label{eq_SM:Az_continuous_L} \\ 
    \partial_xA_z(L^+) - 
    \partial_xA_z(L^-) &= 
    \frac{\Lambda_2}{c'^2}  E_0^y
    \theta(L^-)
    \label{eq_SM:dAz_jump_L}
\end{align}

\subsection{Left-incident photon}

Outside the slab, the photon propagates freely with 
vacuum wavevector $k_0 = \omega$ ($c=1$). For a left-incident 
photon, 
\begin{align}
    A_z(x) = \begin{cases}
        e^{ik_0x} + re^{-ik_0x}, \quad &x < 0, \\
        a_+ e^{i k_+ x} + a_- e^{i k_- x}, \quad & 0 < x < L, \\
        t e^{i k_0 x}, \quad & x>L.
    \end{cases}  
\end{align}
where $r,t$ are the reflection and transmission coefficients, respectively, and $a_\pm$ are unknown amplitudes. Fig. \ref{fig:Transmission}(a) shows the setup for the left-to-right transmission. From 
Eq.~(\ref{eq_SM:theta_slaved}), the associated axion field inside the slab is
\begin{equation}\label{eq_SM:theta_inside}
    \theta(x) = \begin{cases}
        0,  & x < 0 \\\frac{i\Lambda_1
    }{m^2-\omega^2- i \Gamma \omega} \Big[ (E_0^yk_+ - \omega B_0^z)
    a_+e^{ik_+x} + 
    (E_0^yk_- - \omega B_0^z)
    a_-e^{ik_-x} \Big],  & 0<x<L, \\
    0, & x>L.
    \end{cases}
\end{equation}

We now apply the four boundary conditions 
Eqs.~(\ref{eq_SM:Az_continuous}), 
(\ref{eq_SM:dAz_jump}, 
\ref{eq_SM:Az_continuous_L}, 
\ref{eq_SM:dAz_jump_L}) to solve for 
$(a_+, a_-, r, t)$.

\textit{At $x=0$:}

From $A_z$ continuity 
Eq.~(\ref{eq_SM:Az_continuous}),
\begin{equation}\label{eq_SM:LR1}
    a_+ + a_- = 1 + r.
\end{equation}
From the modified $\partial_xA_z$ condition 
Eq.~(\ref{eq_SM:dAz_jump}),
\begin{equation}\label{eq_SM:LR2}
    ik_+a_+ + ik_-a_- - 
    i \frac{\mathcal{E}}{c'^2 (m^2 - \omega^2 - i \Gamma \omega)} \left( \mathcal{E} k_+ - \omega \mathcal{B} \right)  a_+ - i \frac{\mathcal{E}}{c'^2 (m^2 - \omega^2 - i \Gamma \omega)} \left( \mathcal{E} k_- - \omega \mathcal{B} \right)  a_-   
     = ik_0(1-r),
\end{equation}
where $\mathcal{E} = E_0^y \sqrt{\Lambda_1 \Lambda_2}, \mathcal{B} = B_0^z \sqrt{\Lambda_1 \Lambda_2} $. We define
\begin{equation}\label{eq_SM:ktilde}
   \Delta_\pm = \frac{\mathcal{E} k_\pm - \omega \mathcal{B}}{m^2 - \omega^2 - i \Gamma \omega}  ,
\end{equation}
So that Eq.(\ref{eq_SM:LR2}) becomes
\begin{equation}\label{eq_SM:LR2_simple}
    \left( k_+ - \mathcal{E} \Delta_+ /c'^2 \right) a_+ + \left( k_- - \mathcal{E} \Delta_-/c'^2 \right)a_- = ik_0(1-r).
\end{equation}

\textit{At $x=L$:}

From $A_z$ continuity 
Eq.(\ref{eq_SM:Az_continuous_L}), we obtain
\begin{equation}\label{eq_SM:LR3}
    a_+e^{ik_+L} + a_-e^{ik_-L} = te^{ik_0L}.
\end{equation}
From 
Eq.(\ref{eq_SM:dAz_jump_L}), we get
\begin{equation}\label{eq_SM:LR4}
   \left( k_+ - \mathcal{E} \Delta_+/c'^2 \right) a_+ e^{i k_+ L} + \left( k_- - \mathcal{E} \Delta_-/c'^2 \right)a_- e^{i k_- L} = k_0te^{ik_0L}.
\end{equation}
Eqs.(\ref{eq_SM:LR1}, \ref{eq_SM:LR2_simple}, 
\ref{eq_SM:LR3}, \ref{eq_SM:LR4}) form a 
$4\times4$ linear system for $(a_+, a_-, r, t)$ of the form 
\begin{equation}
    \mathcal{A}_{LR} \begin{pmatrix}
        a_+ \\ a_- \\ r \\ t  
    \end{pmatrix} = \textbf{b}_{LR},
\end{equation}
where
\begin{equation}\label{eq_SM:system4x4}
    \mathcal{A}_{LR} =\begin{pmatrix}
        1 & 1 & -1 & 0 \\
        k_+ - \mathcal{E} \Delta_+ /c'^2 & k_- - \mathcal{E} \Delta_- /c'^2& k_0 & 0 \\
        e^{ik_+L} & e^{ik_-L} & 0 & -e^{ik_0L} \\
        \left( k_+ - \mathcal{E} \Delta_+ /c'^2\right)e^{ik_+L} & \left( k_- - \mathcal{E} \Delta_- /c'^2\right) e^{ik_-L} 
        & 0 & -k_0e^{ik_0L}
    \end{pmatrix}, \quad
    \textbf{b}_{LR} = 
    \begin{pmatrix} 1 \\ k_0 \\ 0 \\ 0 
    \end{pmatrix}.
\end{equation}
The transmission coefficient is given by $t_{LR} =  \left( \mathcal{A}_{LR}^{-1} \textbf{b}_{LR} \right)_{4} $

\begin{figure*}
\includegraphics[width=0.8 \textwidth]{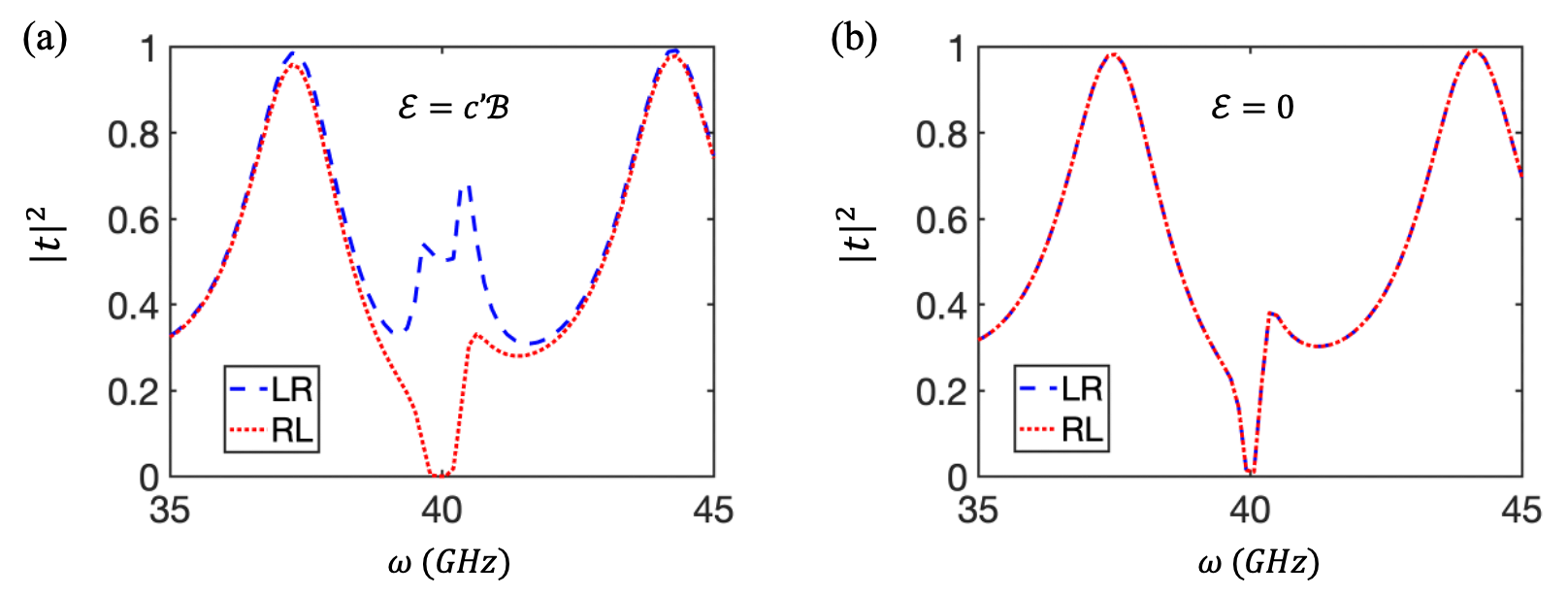}
\caption{ The transmissivities $|t_{LR}|^2$ and $|t_{RL}|^2$ as a function of photon frequency $\omega$ when (a) $\mathcal{E} = c' \mathcal{B}$ and (b) $\mathcal{E} = 0$.
}
\label{fig:Trans}
\end{figure*}

\subsection{Right-incident photon}

Outside the slab, the photon propagates freely with 
vacuum wavevector $k_0 = \omega$. For a right-incident 
photon,
\begin{align}
    A_z(x) = \begin{cases}
        t' e^{-ik_0x}, \quad &x < 0, \\
        a'_+ e^{i k_+ x} + a'_- e^{i k_- x}, \quad & 0 < x < L, \\
        e^{-ik_0x} + r'e^{ik_0x}, \quad & x>L.
    \end{cases}  
\end{align}
where $r', t'$ are the reflection and transmission 
coefficients respectively, and $a'_\pm$ are unknown 
amplitudes. The associated axion field inside the slab 
is, from Eq.~(\ref{eq_SM:theta_slaved}):
\begin{equation}
    \theta(x) = \begin{cases}
        0,  & x < 0 \\\frac{i\Lambda_1
    }{m^2-\omega^2} \Big[ (E_0^yk_+ - \omega B_0^z)
    a'_+e^{ik_+x} + 
    (E_0^yk_- - \omega B_0^z)
    a'_-e^{ik_-x} \Big],  & 0<x<L, \\
    0, & x>L.
    \end{cases}
\end{equation}

We apply the four boundary conditions 
Eqs.~(\ref{eq_SM:Az_continuous}), 
(\ref{eq_SM:dAz_jump}), 
(\ref{eq_SM:Az_continuous_L}), 
(\ref{eq_SM:dAz_jump_L}) to solve for 
$(a'_+, a'_-, r', t')$.

\textit{At $x=0$:}

From $A_z$ continuity 
Eq.~(\ref{eq_SM:Az_continuous}):
\begin{equation}\label{eq_SM:RL1}
    a'_+ + a'_- = t'.
\end{equation}
From the modified $\partial_xA_z$ condition, we have
Eq.~(\ref{eq_SM:dAz_jump}):
\begin{equation}\label{eq_SM:RL2}
    \left( k_+ - \mathcal{E} \Delta_+ /c'^2 \right) a'_+ + \left( k_- - \mathcal{E} \Delta_- /c'^2 \right)a'_- = - k_0t'.
\end{equation}

\textit{At $x=L$:}

From $A_z$ continuity 
Eq.~(\ref{eq_SM:Az_continuous_L}):
\begin{equation}\label{eq_SM:RL3}
     a'_+ e^{ik_+L} + a'_- e^{ik_-L} = 
    e^{-ik_0L} + r'e^{ik_0L}.
\end{equation}
From the modified $\partial_xA_z$ condition Eq.~(\ref{eq_SM:dAz_jump_L}),
\begin{equation}\label{eq_SM:RL4}
    \left( k_+ - \mathcal{E} \Delta_+ /c'^2 \right) a'_+e^{ik_+L} + 
    \left( k_- - \mathcal{E} \Delta_- /c'^2 \right) a'_-e^{ik_-L} = 
    -k_0e^{-ik_0L} + k_0r'e^{ik_0L}.
\end{equation}
Eqs.~(\ref{eq_SM:RL1}, \ref{eq_SM:RL2}, 
\ref{eq_SM:RL3}, \ref{eq_SM:RL4}) form a 
$4\times4$ linear system for $(a'_+, a'_-, r', t')$ 
of the form
\begin{equation}
    \mathcal{A}_{RL} \begin{pmatrix}
        a'_+ \\ a'_- \\ r' \\ t'  
    \end{pmatrix} = \textbf{b}_{RL},
\end{equation}
where
\begin{equation}\label{eq_SM:system4x4_RL}
    \mathcal{A}_{RL} = \begin{pmatrix}
        1 & 1 & 0 & -1 \\
        k_+ - \mathcal{E} \Delta_+/c'^2 & k_- - \mathcal{E} \Delta_-/c'^2 & 0 & k_0 \\
        e^{ik_+L} & e^{ik_-L} & -e^{ik_0L} & 0 \\
        \left(k_+ - \mathcal{E} \Delta_+/c'^2 \right)e^{ik_+L} & \left(k_- - \mathcal{E} \Delta_-/c'^2 \right)e^{ik_-L} 
        & -k_0e^{ik_0L} & 0
    \end{pmatrix}, \quad
    \textbf{b}_{RL} = 
    \begin{pmatrix} 0 \\ 0 \\ e^{-ik_0L} \\ 
    -k_0e^{-ik_0L}
    \end{pmatrix}.
\end{equation}
The transmission coefficient is given by 
$t_{RL} =  
\left(\mathcal{A}_{RL}^{-1}\textbf{b}_{RL}
\right)_{4}$.

\subsection{Numerical results}

We plot the transmissivities $|t_{LR}|^2$ and $|t_{RL}|^2$ in Fig. \ref{fig:Trans}. At perfect nonreciprocity $\mathcal{E} = c' \mathcal{B}$, the transmissivities are unequal near the avoided crossing ($\omega \sim 40$ GHz), as shown in Fig. \ref{fig:Trans}(a). This asymmetry has a transparent physical origin rooted in the nonreciprocal dispersion. At perfect nonreciprocity, the two polariton wavevectors inside the slab are given by Eq.(\ref{eq_SM:kpm2}). Crucially, $k_+ = \omega/c'$ is purely real regardless of the dissipation $\Gamma$. This is a direct consequence of perfect nonreciprocity: the right-moving photon is completely decoupled from the lossy axion mode and propagates freely through the slab without decay, giving high transmission $|t_{LR}|^2$, reduced below unity only by Fresnel reflection at the interfaces due to the mismatch between the photon velocities in vacuum and inside the slab ($c \neq c'$). In contrast, $k_-$ acquires a finite imaginary part due to $\Gamma$, since the left-moving photon hybridizes strongly with the dissipative axion mode, leading to the axion absorbing energy, causing exponential attenuation of this mode inside the slab. When $\mathcal{E}=0$, the 
wavevectors satisfy $k_+ = -k_-$ (Eq. (\ref{eq_SM:kpme0})), restoring reciprocity, and the transmissivities are equal, as shown in Fig. \ref{fig:Trans}(b). We note that the polaritonic gap --- defined as the frequency range $m < \omega < \sqrt{m^2 + \mathcal{B}^2
}$ in which $k_\pm$ are purely imaginary --- is precisely defined only in the dissipationless limit $\Gamma = 0$ (Eq.\ref{eq_SM:kpme0gamma0}). For small but finite $\Gamma$, the modes acquire small real parts inside the gap and are no longer purely evanescent. Nevertheless, their imaginary parts remain large and the transmission is strongly suppressed near $\omega \sim 40$ GHz, as seen in Fig. \ref{fig:Trans}(b).


\end{document}